# Exact elegant Laguerre-Gaussian vector wave packets

W. Nasalski

*Institute of Fundamental Technological Research, Polish Academy of Sciences, A. Pawińskiego 5b, 02-106 Warsaw, Poland*
*wnasal@ippt.gov.pl*

An exact closed-form representation is derived of a vector elegant Laguerre-Gaussian wave packet. Its space-time representation consists of three mutually orthogonal field components - of a common azimuthal index and different radial indices - uniquely distinguished by first three powers of the paraxial parameter. The transverse components are of tm-radial and te-azimuthal polarization and appear, under their normal incidence, to be eigenmodes of any horizontally planar, homogeneous and isotropic structure, with eigenvalues given by the reflection and transmission coefficients. In this context, the interrelations between the cross-polarization symmetries of wave packets in free space and at medium planar interfaces are discussed.

Elegant (complex-argument) Hermite-Gaussian (EHG) and Laguerre-Gaussian (ELG) paraxial beams were introduced in optics a few decades ago.[1-3] Since then, owing to their attractive analytical properties, they have been widely used in description of spatial structures of optical field. The recent interest in the ELG beams stems to a large extent from their property of possessing well defined angular momentum (AM).[4,5] In addition, several contributions to the topic of nonparaxial ELG beams have been also reported.[5-11] This report will present an alternative nonparaxial solution, which follows directly basic features of the ELG beams known from their paraxial representation.[12]

The exact representation of the ELG field, depending both on space coordinates and time, will be derived starting from two observations. First, characteristic symmetries of paraxial ELG beams indicate that their exact versions should exhibit symmetries of the same type. Secondly, Gaussian beams, commonly regarded as paraxial solutions to the Helmhotz equation, also yield exact solutions to the d'Alembert wave equation. The notion of the packetlike solutions of this type has been used, for example, in inventions of focus wave modes.[13-17] Their extension to higher-order ELG scalar solutions was also proposed, restricted however to fields of axial symmetry.[14] The solution of this type, with vortex terms already included, was promoted recently[10] within the field complex vector approach,[18] although with no reference to focus wave modes. The analysis was given in the spectral domain by use of Bessel functions. The analysis here is different. It is given directly in the space-time domain by use of ELG functions. It follows the lines of Ref. 12 on paraxial EHG and ELG monochromatic beams.

Consider the electromagnetic monochromatic field in an isotropic, homogeneous and unbounded medium, defined by its intrinsic impedance Z and admittance Y. The field is assumed in a beamlike form, with its wave number $k$ and spectral wave vectors $\underline{k}=(k_x, k_y, k_z) = (\underline{k}_\perp, k_z)$, the propagation direction along the $z$-axis, the transverse $w_0$ and longitudinal $z_D = kw_0^2$ scales and the parameter $f = 2^{-1/2} w_0/z_D$ indicating the field paraxiality level. The field is considered in the circular polarization frame $(\hat{\underline{e}}_R, \hat{\underline{e}}_L, \hat{\underline{e}}_z)$. The coordinates $\varsigma_\pm$ and $z_\pm$, are defined as normed by $w_0$ and $z_D$, respectively: $\varsigma_\pm w_0 = 2^{-1/2}(x \pm iy)$ and $z_\pm z_D = z \pm ct$. The electric $\underline{E}$ and magnetic $\underline{H}$ field vectors, normed by $Y^{1/2}$ and $Z^{1/2}$, respectively, are expressed by vector Hertz potentials $\underline{M}$ and $\underline{N}$ satisfying the wave equation.[19] They are normed here by $w_0^2$. The transverse - with respect to the $z$-axis - magnetic (tm) optical field is generated by the $z$-component $w_0^2 M_z$ of the potential $\underline{M}$:

$$\underline{E}^{(tm)} = 2^{1/2} f (\hat{\underline{e}}_R \partial_z \partial_{\varsigma_+} + \hat{\underline{e}}_L \partial_z \partial_{\varsigma_-}) M_z - 2\hat{\underline{e}}_z \partial_{\varsigma_+} \partial_{\varsigma_-} M_z, \quad (1)$$

$$\underline{H}^{(tm)} = i 2^{1/2} f (\hat{\underline{e}}_R \partial_{ct} \partial_{\varsigma_+} - \hat{\underline{e}}_L \partial_{ct} \partial_{\varsigma_-}) M_z, \quad (2)$$

The dual transverse electric (te) field can be obtained by the replacement $-\underline{H}^{(tm)}$ by $\underline{E}^{(te)}$ and $\underline{E}^{(tm)}$ by $\underline{H}^{(te)}$ for $\underline{N} = \underline{M}$. Next, the potential $M_z = G \exp(i2^{-1} f^{-2} z_-)$ is decomposed into two factors: the propagation factor $\exp(i2^{-1} f^{-2} z_-)$ dependent on $z_-$ and the modulation function $G \equiv G(\varsigma_+, \varsigma_-, z_+)$ dependent on $z_+$. Then, the wave equation converts exactly to the Fock equation:

$$(2i\partial_{z_+} + \partial_{\varsigma_+} \partial_{\varsigma_-}) G_{p,\pm l} = 0; \qquad G_{p,\pm l} = \partial_{\varsigma_\pm}^p \partial_{\varsigma_\mp}^{p+l} g \quad (3)$$

with the fundamental Gaussian solution for $p = 0 = l$: $g = v^{-2} \exp(-u^2)$, $u^2 = \varsigma_+ \varsigma_- v^{-2}$, of the beam complex, real and phase front radii: $v$, $w$ and $R$, respectively; $v^{-2} = (1 + iz_+/2)^{-1} = w^{-2} - iz_D R^{-1}$ is normed by $w_0^{-2}$.[12] The differential operations on $g$ in Eq. (3) yield the unnormed higher-order ELG modes $G \equiv G_{p,\pm l}$ of the radial $p$ and azimuthal $\pm l$ mode indices p, $l \geq 0$[12,20,21] and, by the definition of the associated Laguerre polynomials $[(d/du^2 u^2)^p - p! L_p^l(u^2)] u^{2l} \exp(-u^2) = 0$, the ELG field representation in the configuration (Eq. (4)) and spectral (Eq. (5)) domains, respectively:

$$G_{p,\pm l} = (-1)^{p+l} v^{-2(p+l+1)} \varsigma_\pm^l p! L_p^l(u^2) \exp(-u^2), \quad (4)$$

$$\tilde{G}_{p,\pm l} = i^{2p+l} \kappa_\mp^p \kappa_\pm^{p+l} \exp(-\kappa_+ \kappa_- v^2), \quad (5)$$

where $\kappa_\pm w_0^{-1} = 2^{-1/2}(k_x \pm i k_y)$, $\varsigma_\pm = vu \exp(\pm i\phi)$ and $\phi$ is the spatial azimuthal angle (cf. Eqs. (48)-(58) in Ref. 12). The exact fundamental Gaussian appears still in the standard form, well known from paraxial analyses, in spite of the length of its longitudinal scale $z_D' = 2^{-1} z_D$ shortened here by two. However, the beam complex radius $v$ depends now on $z_+$ instead of $z$, what causes the evolution of the field in Eqs. (4)-(5) in time through changes of its waist position always placed at $z = -ct$. At any time moment $t$ or at the beam phase front $z_+ = 2z$ the exact Gaussian copies the standard Gaussian, with its waist placed at $z = -ct$ or at $-z$, respectively. In such cases, all orders of the ELG functions defined in Eqs. (4)-(5) represent the spatial beams with the diffraction length $z_D'$. The solution describes exactly the field spatial structure at any time moment and, at any transverse plane $z = const.$, the time evolution of the field. Inserting Eqs. (4) into Eqs. (1)-(2) yields, after dropping the factor $\exp(i2^{-1}f^{-2}z_-)$:

$$\underline{E}^{(tm)} = i2^{-1/2}\hat{\underline{e}}_{\{R,L\}}(f^{-1}G_{p+1,\pm(l-1)} + f^{+1}G_{p+2,\pm(l-1)}) \\ + i2^{-1/2}\hat{\underline{e}}_{\{L,R\}}(f^{-1}G_{p,\pm(l+1)} + f^{+1}G_{p+1,\pm(l+1)}) - 2\hat{\underline{e}}_z G_{p+1,\pm l}, \quad (6)$$

$$\underline{H}^{(tm)} = \pm 2^{-1/2}\hat{\underline{e}}_{\{R,L\}}(f^{-1}G_{p+1,\pm(l-1)} - f^{+1}G_{p+2,\pm(l-1)}) \\ \mp 2^{-1/2}\hat{\underline{e}}_{\{L,R\}}(f^{-1}G_{p,\pm(l+1)} - f^{+1}G_{p+1,\pm(l+1)}), \quad (7)$$

for tm polarisation and, by duality, for te polarisation: $E^{(te)} = -H^{(tm)}$ and $H^{(te)} = E^{(tm)}$. Note the notation used through the text: a and b in {a,b} refer to the upper or lower signs in respective expressions, for example to $\pm l$ in Eqs. (6)-(7). Further derivation will be restricted only for positive azimuthal indices. The opposite case can be considered per analogy.

As was pointed out in Ref. 12, the transverse field components, here separately for the paraxial (with $f^{-1}$) and nonparaxial (with $f$) terms in Eqs. (6)-(7), display the clear symmetry $(\hat{\underline{e}}_R, p+1, l-1) \leftrightarrow (\hat{\underline{e}}_L, p, l+1)$ - the R circular field components are associated with radial indices greater by one and azimuthal indices less by two than the corresponding indices of the L circular field components. This cross-polarization symmetry (XPS) is exactly given by the two ELG relations:[12]

$$G_{p,l\mp 1}\exp(\pm 2i\phi) = G_{p\mp 1,l\pm 1}; \\ G_{p,l\mp 1}\exp(\pm i\phi) = G_{p\mp 1/2,l}. \quad (8)$$

Another symmetry $(\hat{\underline{e}}_R, -\phi) \leftrightarrow (\hat{\underline{e}}_L, +\phi)$ appears from the definitions $\hat{\underline{e}}_\rho \pm i\hat{\underline{e}}_\phi = 2^{1/2}\hat{\underline{e}}_{\{R,L\}}\exp(\mp i\phi)$ of radial and azimuthal polarization in the cylindrical basis $(\hat{e}_\rho, \hat{e}_\phi, \hat{e}_z)$. Owing to all these symmetries together, Eqs. (6)-(7) can be concisely restated as:

$$\underline{E}^{(tm)} = i\hat{\underline{e}}_\rho (f^{-1}G_{p+1/2,l} + f^{+1}G_{p+3/2,l}) - 2\hat{\underline{e}}_z G_{p+1,l}, \quad (9)$$

$$\underline{H}^{(tm)} = i\hat{\underline{e}}_\phi (f^{-1}G_{p+1/2,l} - f^{+1}G_{p+3/2,l}), \quad (10)$$

and the total field of the vector wave packet, with the complex amplitudes $\alpha$ and $\beta$ of the tm-radial (for $\beta = 0$) and te-azimuthal (for $\alpha = 0$) polarization, reads:

$$\underline{E} = if^{-1}(\alpha\hat{\underline{e}}_\rho - \beta\hat{\underline{e}}_\phi)G_{p+1/2,l} - 2\alpha\hat{\underline{e}}_z G_{p+1,l} \\ + if^{+1}(\alpha\hat{\underline{e}}_\rho + \beta\hat{\underline{e}}_\phi)G_{p+3/2,l}, \quad (11)$$

$$\underline{H} = if^{-1}(\beta\hat{\underline{e}}_\rho + \alpha\hat{\underline{e}}_\phi)G_{p+1/2,l} - 2\beta\hat{\underline{e}}_z G_{p+1,l} \\ + if^{+1}(\beta\hat{\underline{e}}_\rho - \alpha\hat{\underline{e}}_\phi)G_{p+3/2,l}. \quad (12)$$

The representation shown in Eqs. (11)-(12) is the exact version of the paraxial ELG beams,[12] straightforward, as obtained directly from the definitions in Eq. (3) and the symmetry relations in Eq. (8), and complete, as given, by Eqs. (4) and (5), in both the configuration and spectral domains. It is also analytic, as expressed by elements of the complete set of square-integrable, non-singular ELG functions. The R and L circular field components are obtained for $\beta = \mp i\alpha$. The three parts of the solution are labelled separately by the powers of $f$: $f^{\mp 1}$ and $f^0$, for the paraxial, nonparaxial and longitudinal terms and enter the exact solution symmetrically with respect to $l$ and with different values of the radial index $p+1/2$, $p+3/2$ and $p+1$, respectively. In the paraxial ($f \ll f^{-1}$) and nonparaxial ($f \gg f^{-1}$) ranges, however, the field is already of common polarization and distribution:

$$\underline{E} \cong if^{\mp 1}(\alpha\hat{\underline{e}}_\rho \mp \beta\hat{\underline{e}}_\phi)G_{\{p+1/2,p+3/2\},l}, \quad (13)$$

$$\underline{H} \cong if^{\mp 1}(\beta\hat{\underline{e}}_\rho \pm \alpha\hat{\underline{e}}_\phi)G_{\{p+1/2,p+3/2\},l}. \quad (14)$$

Note that the te and tm vector beams, still of axial symmetry, also exist in the paraxial range, expressed by the first-order Bessel functions of the first kind.[22]

The final form of the ELG wave packet is obtained by specification in Eqs. (11)-(12) the indices $p$ and $l$ and the complex amplitudes $\alpha$ and $\beta$ or, in general, by summing up through the set of all these parameters:

$$\{\underline{E},\underline{H}\}(z_+,\omega t) = \sum_{\alpha,\beta}\sum_{p,l}\{\underline{E},\underline{H}\}(z_+)\exp(i\omega z_-/c), \quad (15)$$

where $\omega = kcz_D$. The representation (15) is exact as exact are all of its field components, but still it is only the packetlike ELG solution. For its fully polychromatic version, the spectral weight-function $\psi(\omega)$ of a positive support should be defined first and applied in (15) for the pulse composition. That, for the constant $z_D$, yields the Fourier-Laplace transform of the wave packet field:

$$\{\underline{\hat{E}},\underline{\hat{H}}\}(z_+,t) = \pi^{-1}\int\{\underline{E},\underline{H}\}(z_+,\omega t)\psi(\omega)d\omega. \quad (16)$$

With proper integrability conditions for $\psi(\omega)$ the pulses of finite-energy can be then obtained. Note also that the second family of the exact ELG wave packets Eqs. (11)-(12) is obtained by the exchange $z_+ \leftrightarrow z_-$ in the Eqs. (1)-(5). For $p = 0 = l$ such solutions are equivalent to the Gaussian focus modes[13-17] used previously in synthesis of finite-energy pulses.[16] Which version of the placement of $z_\pm$ in Eqs. (1)-(5) is more suitable in such analysis is the matter of further examination.

The symmetries of the ELG beams are particularly vivid under their normal incidence and the cross-polarisation coupling (XPC) at any planar structure composed, in general, of several homogeneous, lossless, and isotropic layers. That was theoretically analysed and numerically verified in Ref. 12 for $kw_0 = 2\pi$, that is close to the paraxial limit. For exact wave packets the reflection and transmission matrices are of the same form, written here in the basis $(\hat{\underline{e}}_R, \hat{\underline{e}}_L, \hat{\underline{e}}_z)$ at the upper ($z = z_a$) and lower ($z = z_b$) planes of the structure:[12]

$$\underline{\underline{r}} = \begin{bmatrix} r_+ e^{+2i\varphi} & r_- \\ r_- & r_+ e^{-2i\varphi} \end{bmatrix}; \qquad \underline{\underline{t}} = \begin{bmatrix} t_+ & t_- e^{-2i\varphi} \\ t_- e^{+2i\varphi} & t_+ \end{bmatrix}. \quad (17)$$

The matrices (17) relate the parallel to the interfaces spectral components of the incident, reflected and transmitted fields. Their elements $r_\pm = 2^{-1}(r_p \pm r_s)$ and $t_\pm = 2^{-1}(\eta t_p \pm t_s)$ are expressed by the sums and differences of the TM (p) and TE (s) transmission $\eta t_p, t_s$ and reflection $r_p, r_s$ coefficients.[12] In the case of a single interface they are just the standard Fresnel coefficients for $\eta = 1$. The parameter $\eta$ stands for the ratio of the direction cosines of the transmitted and incident waves and $\varphi$ is the spectral azimutal angle. By applying Eqs. (17) and (8) to Eqs. (11)-(12), the two eigenvalue equations for the tm field and the two for the te field are obtained:

$$\underline{\underline{r}} \widetilde{\underline{E}}_\perp^{(tm)} = r_p \widetilde{\underline{E}}_\perp^{'(tm)}; \qquad \underline{\underline{t}} \widetilde{\underline{E}}_\perp^{(tm)} = \eta t_p \widetilde{\underline{E}}_\perp^{(tm)}, \quad (18)$$

$$\underline{\underline{r}} \widetilde{\underline{E}}_\perp^{(te)} = -r_s \widetilde{\underline{E}}_\perp^{'(te)}; \qquad \underline{\underline{t}} \widetilde{\underline{E}}_\perp^{(te)} = t_s \widetilde{\underline{E}}_\perp^{(te)}, \quad (19)$$

with the components $\widetilde{\underline{E}}_\perp = (\widetilde{E}_R, \widetilde{E}_L)$ and $\widetilde{\underline{E}}_\perp' = (\widetilde{E}_L, \widetilde{E}_R)$ transverse with respect to the normal to the structure. Eqs. (18)-(19) are the exact versions of the normal modes defined in Ref. 12 for paraxial beams. The tm-radial and te-azimuthal spectral components of the vector ELG wave packets appear to be eigenvectors of the matrices in Eqs. (17), with eigenvalues equal to p and s reflection and transmission coefficients of the structure. These matrices convert the R field components to the L field components and vice versa, and change their complex amplitudes according to Eqs. (8) and Eqs. (18)-(19). In effect, at $z = z_a, z_b$, the normally incident tm or te ELG fields do not change their polarization and mode parameters under the XPC interactions and the total AM of them, as well as of their paraxial and nonparaxial parts *separately*,[12] is conserved at the interfaces defined above.

The field components are always interrelated in the 3D vector beams or wave packets independently of any approximation, incidence or field type chosen. The XPS relations in Eq. (8), describing the field symmetries are parallel to the XPC relations in Eq. (17), describing the field coupling at planar structures. These symmetries are usually accounted for by the polarization parameters, like $\widetilde{\chi}_{x,y} = \widetilde{E}_x/\widetilde{E}_y$ or $\widetilde{\chi}_{R,L} = \widetilde{E}_R/\widetilde{E}_L$. The matrices in Eq. (17) can be made diagonal by these parameters, with their elements equal, for example, $t_+ + t_- \widetilde{\chi}_{R,L}^{\mp 1} \exp(\mp i\varphi)$ for transmission (cf. Eqs. (11)-(13) and (19) in Ref. 12). The XPC may be not substantial but always is non-negligible and significant for reasons given above, even at dielectric interfaces. It can be considerably increased at interfaces of other types or structures composed of several such interfaces, especially near their resonances. An almost 100% XPC polarization switch of paraxial EHG beam can be achieved, for example, in the metamaterial open cavity.[23]

It would be interesting to discuss the other results based, for example, on field expansions, complex ource/sink models, Bessel beams or focus wave modes,[5-11,13-17,22] in the context of the solution given by Eqs. (4)-(12) and its symmetries in Eqs. (8) and (17). However, such the analysis is outside the scope of this paper. Note only that the expressions for ELG solutions presented in Ref. 12 and here can be readily translated into vector Bessel fields by the standard Fourier-Bessel integral transforms.[19] The spectral parts of these solutions, as defined by Eq. (51) in Ref. 12 and Eq. (5) here, can be directly expressed by Bessel functions as well.

For what is believed the solution derived here is new. It yields the complete, analytic, exact description of the vector ELG wave packets. Similar analysis also can be applied to the EHG wave packets. The formalism presented is so transparent and straightforward that it may serve as a convenient tool in treatments of several problems on optical beams and pulses and their interactions with matter.